\documentclass[prb,twocolumn,floatfix]{revtex4-1}
\usepackage{epsfig}
\usepackage{color}
\usepackage{subfigure}
\usepackage{mathrsfs}
\usepackage{amsmath}
\usepackage{amssymb}
\usepackage{hyperref}
\usepackage{enumerate}
\usepackage{braket}
\usepackage{comment}
\usepackage{yfonts}
\hypersetup{colorlinks=true}
\usepackage{bm}

\newcommand{\be}{\begin{equation}}
\newcommand{\ee}{\end{equation}}
\newcommand{\ba}{\begin{aligned}}
\newcommand{\ea}{\end{aligned}}

\def\H1{\widehat{H}_1}

\newcommand{\bea}{\begin{eqnarray}}
\newcommand{\eea}{\end{eqnarray}}

\def\nn{\nonumber\\}
\def\fr#1{(\ref{#1})}

\def\sgn{{\rm sgn}}

\begin{document}
\title{Integrable spin chains with random interactions}
\author{Fabian H.L. Essler$^1$, Rianne van den Berg$^2$ and Vladimir Gritsev$^{2,3}$}
\affiliation{\mbox{$^1$The Rudolf Peierls Centre for Theoretical
    Physics, Oxford University, Oxford, OX1 3NP, United
    Kingdom}\\$^2$Institute for Theoretical Physics, Universiteit van
  Amsterdam, Science Park 904, 
Postbus 94485, 1098 XH Amsterdam, The Netherlands\\
$^3$Russian Quantum Center, Skolkovo, Moscow 143025, Russia}
\date{\today}
\begin{abstract}
We study a Yang-Baxter integrable quantum spin-1/2 chain with
random interactions. The Hamiltonian is local and involves two and
three-spin interactions with random parameters. We show that
the energy eigenstates of the model are never localized and in fact
exhibit perfect energy and spin transport at both zero and infinite
temperatures. By considering the vicinity of a free fermion point in
the model we demonstrate that this behavior persists under
deformations that break Yang-Baxter integrability but preserve the 
free fermion nature of the Hamiltonian. In this case the ballistic
behavior can be understood as arising from the correlated nature of
the disorder in the model. We conjecture that the model belongs to a
broad class of models avoiding localization in 1D. 
\end{abstract}
\maketitle
\section{Introduction}
After the fundamental paper by Anderson \cite{PWA} it was believed for
a long time that in one-dimensional random potentials all eigenstates
are localized in the thermodynamic limit for arbitrarily weak disorder
\cite{mott,ishii,efetov}. If Azbel resonances \cite{azbel,azbel2}, which form
a set of measure zero, are neglected, the above statement is
rigorously speaking valid only for white noise spatially uncorrelated
randomness \cite{pastur}. Later it was realized 
that the spatial correlations of the disorder potential can have
profoundly influence Anderson localization \cite{JK,F,KM}. 
In this case localization can be partially suppressed, at 
least for weak disorder \cite{IK}. In this context a
delocalization-localization transition in 1D for {\it long-range} 
correlated disorder potentials has been intensively discussed
in the literature \cite{ML1,ML2,ML3,ML4,stanley1,stanley2}. On the other hand, it was
found that models with specific short-range correlated disorder,
so-called dimer models, exhibit conducting states \cite{dimer1,dimer2,dimer3,dimer4}. 
In recent years considerable efforts have been made to understand
the combined effects of disorder and interactions, which leads to the
phenomenon of \emph{many-body localization} (MBL)
\cite{BAA,MBL1,MBL3,MBL4,MBL5,MBL6,MBL7,MBL8}, see
Refs~\cite{MBLR1,MBLR2,MBLR3,AP,PV} for recent reviews.
The MBL transition generally occurs at finite energy densities and is
characterized by ergodicity breaking, the existence of an extensive
number of quasi-local integrals of motion in the localized
phase \cite{MBL7,MBL8,MBL9} and Poissonian level statistics. This is
reminiscent of Yang-Baxter integrable many-body
systems \cite{Korepinbook,Gaudinbook}, which also feature Poissonian level
statistics and extensive numbers of conservation laws. In Yang-Baxter
integrable systems the conserved charges are extensive but have
(quasi) local densities. An interesting question is then whether there
are any connections between Yang-Baxter integrability and MBL. An
example of a Yang-Baxter integrable model that is localized is
provided by disordered Richardson models \cite{BDS}. However, this
class of models is infinite-ranged whereas studies of MBL have focussed
on models like the spin-1/2 Heisenberg chain with a random white-noise
correlated magnetic field. Other forms of disorder such as
random exchange interactions \cite{DM,DSF} have been explored as well 
\cite{VA,VFPP,nonabel2}, and there appears to be a wide spread belief
that MBL behaviour is a rather generic phenomenon in the strong
disorder regime. Non-MBL behaviour has been found in a disordered
Hubbard chain \cite{PBZ}, but this could be related to the presence of
non-abelian symmetries \cite{nonabel,nonabel2}. Another little
explored issue is what effects correlations in the disorder have on
MBL \cite{hoyos1,MBLcorr2,hoyos2}.

In this work we study a Yang-Baxter integrable model of a
Heisenberg-like spin chain with tuneable randomness and abelian
symmetry. We employ a number of standard tools used to probe for
(many-body) localized 
behaviour: inverse participation ratios, local quantum quench dynamics
and transport properties in energy eigenstates. All methods point to
the same conclusion: the model does not exhibit any traces of
localization irrespective of the magnitudes of the interactions and
disorder. On the contrary, we find that the model is an ideal
conductor for both spin and energy.
Moreover, we show in a
non-interacting limit that by deforming the model by tuning the
correlations between the random interaction parameters (the resulting
model is no longer Yang-Baxter integrable) it is possible to induce
localization. This suggests that the model we study here is a
particular example of a broader class of strongly disordered models in
one dimensions that do not exhibit MBL. 
\begin{figure}[ht]
\begin{center}
\includegraphics[width=0.45\textwidth]{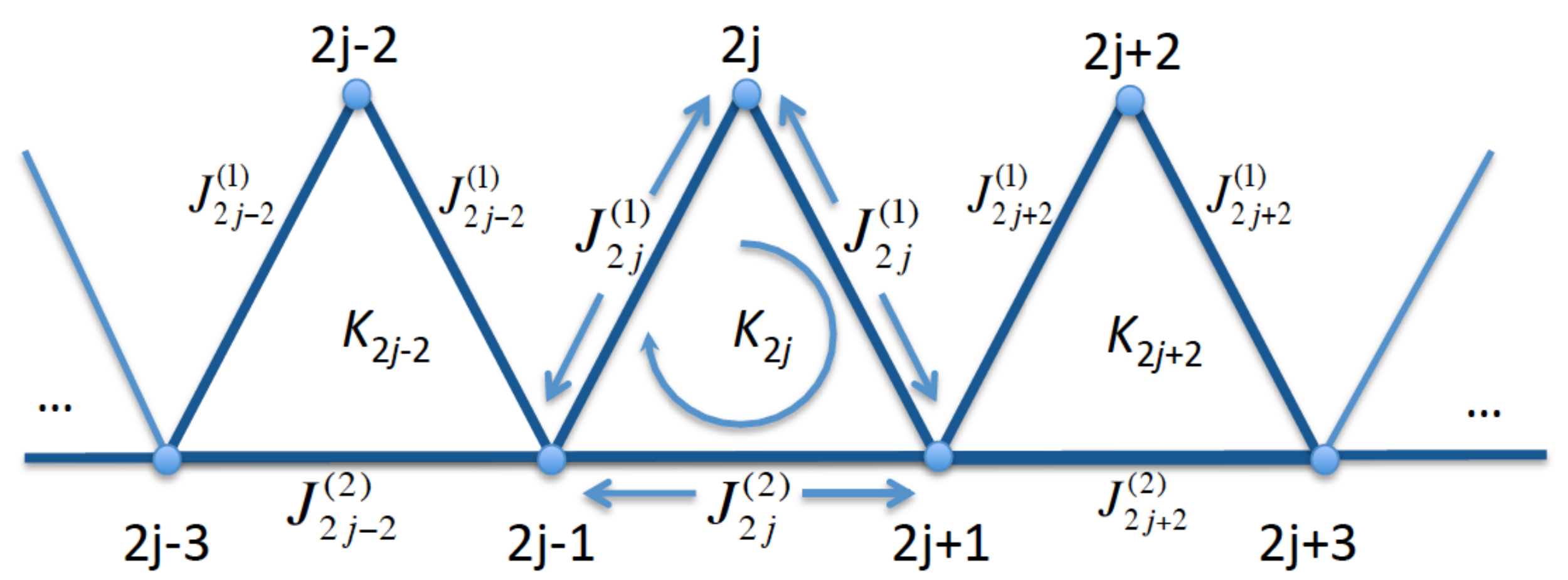}
\caption{Schematic representation of the disordered interacting spin
chain studied here. Three types of position-dependent interactions
with random parameters are present: a nearest-neighbour exchange
$J^{(1)}_{2j}$, a next-nearest neighbour coupling $J^{(2)}_{2j}$ and a 
three-spin interaction $K_{2j}$. Explicit expressions for the various terms
in the Hamiltonian are given in the text. The ratios
$J^{(1)}_{2j}/J^{(2)}_{2j}$ and $J^{(1)}_{2j}/K_{2j}$ are correlated.}
\label{fig:system}
\end{center}
\end{figure}

\section{The Model}
The Hamiltonian of our integrable chain contains nearest-neighbour,
next-nearest neighbour and three-spin interactions with random
couplings, \emph{cf.} Fig. (\ref{fig:system}), and
can be expressed in the form
\bea
H&=&\sum_{j=1}^{L/2}
J^{(1)}_{2j}\Big(
\left[\vec{\sigma}_{2j-1}\cdot\vec{\sigma}_{2j}\right]_{\Delta_{2j}}
+\left[\vec{\sigma}_{2j}\cdot\vec{\sigma}_{2j+1}\right]_{\Delta_{2j}}\Big)\nn
&&\quad+K_{2j}\Big(\left[\vec{\sigma}_{2j}\cdot\big(
\vec{\sigma}_{2j-1}\times\vec{\sigma}_{2j+1}\big)\right]_{\Delta_{2j}^{-1}}+\Delta^{-1}_{2j}
\Big)\nn
&&\quad+J^{(2)}_{2j}\big(\vec{\sigma}_{2j-1}\cdot\vec{\sigma}_{2j+1}-1\big)\ ,
\label{Hamil}
\eea
where $[\vec{\sigma}_j\cdot\vec{\sigma}_k]_\Delta=
\sigma^x_j\sigma^x_k+\sigma^y_j\sigma^y_k+\Delta
(\sigma^z_j\sigma^z_k-1)$. The exchange couplings are parametrized as
\bea
J^{(1)}_{2j}&=&\frac{\sin^2\eta\cosh\xi_{2j}}{\sin^2\eta+\sinh^2\xi_{2j}}\ ,\
J^{(2)}_{2j}=\frac{\cos\eta\sinh^2\xi_{2j}}{\sin^2\eta+\sinh^2\xi_{2j}}\ ,\nn
K_{2j}&=&\frac{\sin\eta\cos\eta\sinh\xi_{2j}}{\sin^2\eta+\sinh^2\xi_{2j}}\ ,\
\Delta_{2j}=\frac{\cos\eta}{\cosh\xi_{2j}}\ ,
\eea
where $\xi_{2j}$ and $\Delta=\cos(\eta)$ are free parameters of the
model. By construction we recover the spin-1/2 Heisenberg XXZ Hamiltonian if
we set all inhomogeneities to zero $\xi_{2}=\xi_4=\dots=\xi_L=0$.
In the following we mainly consider the case where $\xi_{2k}$
are independent random variables drawn from a flat
distribution
\be
P_W(\xi)=\frac{1}{2W}\theta(W-|\xi|)\ .
\label{PD1}
\ee
The derivation of the Hamiltonian \fr{Hamil} is summarized in
Appendix \ref{app:inhom}. The model \fr{Hamil} is a variant of a class
of disordered impurity models previously studied by Kl\"umper and Zvyagin
\cite{KZ1}, who in particular determined thermodynamic properties
\cite{KZ1,KZ2,Z1,Z11,Z2}. Yang-Baxter integrability imposes severe
restrictions on the  form of the Hamiltonian. This results in all
three kinds of interactions involving the same random parameters and
can be viewed as short-range correlated disorder in a model of
interacting spins. 

As a sufficiently strong next-nearest neighbour exchange can induce
dimerization our model can in some sense be considered as an
interacting analogue of the ``dimer models'' mentioned
above. 
\subsection{Higher conservation laws}
\label{sec:HCL}
As shown in Appendix \ref{app:inhom} the Hamiltonian \fr{Hamil} is
related to the transfer matrix $\tau(\mu)$ of an inhomogeneous
six-vertex model. This connection is useful for constructing higher
conservation laws, which are a characteristic feature of Yang-Baxter
integrable models. In the case at hand they
can be obtained by taking logarithmic derivatives of the transfer
matrix at $\mu=0$
\be
Q^{(n)}=i^{n}\frac{d^{n-1}}{d\mu^{n-1}}\Bigg|_{\mu=0}\ln\big(\tau(\mu)\big)\ ,\quad
n=2,3,\dots
\label{HCL}
\ee
The Hamiltonian is by construction proportional to $Q^{(2)}$
\be
H=-2i\sin\eta\ Q^{(2)}\ .
\ee
Importantly the higher conservation laws are also (ultra)local in the
following sense: they can be expressed in the form
\be
Q^{(n)}=\sum_{j=1}^L Q^{(n)}_j\ ,
\ee
where $Q^{(n)}_j$ act non-trivially only on a finite number of
neighbouring sites. We note that the structure of these conservation
laws is very different from that on the ``l-bits'' in many-body
localized systems. 

In the following we will make use of the first higher conservation law
$Q^{(3)}$. To that end we require an explicit expression in terms of
the L-operator \fr{Lop} and its derivatives. For the operator
$Q^{(2)}$ this is readily done:
\be
Q^{(2)}=-\sum_{j=1}^{L/2}Q^{(2,1)}_{2j-1,2j}+Q^{(2,2)}_{2j-1,2j,2j+1}\ ,
\ee
where
\bea
\Big[Q^{(2,2)}_{1,2,3}\Big]_{\alpha_1\alpha_2\alpha_3}^{\beta_1\beta_2\beta_3}
&=&\Big[L(-x_2)\Big]^{\alpha_1c}_{\alpha_2d}
\Big[L'(0)\Big]^{c\beta_3}_{\alpha_3e}
\Big[L(x_2)\Big]^{e\beta_1}_{d\beta_2}\ ,\nn
\Big[Q^{(2,1)}_{1,2}\Big]_{\alpha_1\alpha_2}^{\beta_1\beta_2}
&=&\Big[L'(-x_{2})\Big]^{\alpha_{1}c}_{\alpha_{2}d}
\Big[L(x_{2})\Big]^{c\beta_{1}}_{d\beta_{2}}\ .
\eea
The conservation law $Q^{(3)}$ can be expressed as a sum of terms that
involve spin interactions on two, three, four and five neighbouring
sites respectively
\bea
Q^{(3)}&=&-i\sum_{j=1}^{L/2}\Bigg[ Q^{(3,1)}_{2j-1,2j}+Q^{(3,2)}_{2j-1,2j,2j+1}\nn
&+&Q^{(3,3)}_{2j-1,2j,2j+1,2j+2}
+Q^{(3,4)}_{2j-1,2j,2j+1,2j+2,2j+3}\Bigg],\nn
\label{Q3}
\eea
where
\bea
\Big(Q^{(3,1)}_{1,2}\Big)_{\alpha_1\alpha_2}^{\beta_1\beta_2}
&=&\Big[L''(-x_{2})\Big]^{\alpha_{1}c}_{\alpha_{2}d}
\Big[L(x_{2})\Big]^{c\beta_{1}}_{d\beta_{2}}\nn
&&-\Big[Q^{(2,1)}_{1,2}\ Q^{(2,1)}_{1,2}\Big]_{\alpha_{1}\alpha_{2}}^{\beta_{1}\beta_{2}}\ ,\nn
\Big(Q^{(3,2)}_{1,2,3}\Big)_{\alpha_1\alpha_2\alpha_3}^{\beta_1\beta_2\beta_3}&=&
2\Big[L'(-x_2)\Big]^{\alpha_1c}_{\alpha_2d}
\Big[L'(0)\Big]^{c\beta_3}_{\alpha_3e}
\Big[L(x_2)\Big]^{e\beta_1}_{d\beta_2}\nn
&&-
\Big[Q^{(2,1)}_{1,2}Q^{(2,2)}_{1,2,3}+
Q^{(2,2)}_{1,2,3}Q^{(2,1)}_{1,2}\Big]_{\alpha_1\alpha_2\alpha_3}^{\beta_1\beta_2\beta_3}\ ,\nn
Q^{(3,3)}_{1,2,3,4}&=&
Q^{(2,1)}_{3,4}\ Q^{(2,2)}_{1,2,3}-Q^{(2,2)}_{1,2,3}\ Q^{(2,1)}_{3,4}\ ,\nn
Q^{(3,4)}_{1,2,3,4,5}&=&Q^{(2,2)}_{3,4,5}\ Q^{(2,2)}_{1,2,3}-
Q^{(2,2)}_{1,2,3}\ Q^{(2,2)}_{3,4,5}\ .
\label{Qnm}
\eea
The operator $Q^{(3)}$ can be expressed in terms of Pauli matrices
using \fr{Lop}, but this is not particularly useful for our purposes.


\section{Non-interacting limit}
\begin{figure}[ht]
\begin{center}
\includegraphics[width=0.45\textwidth]{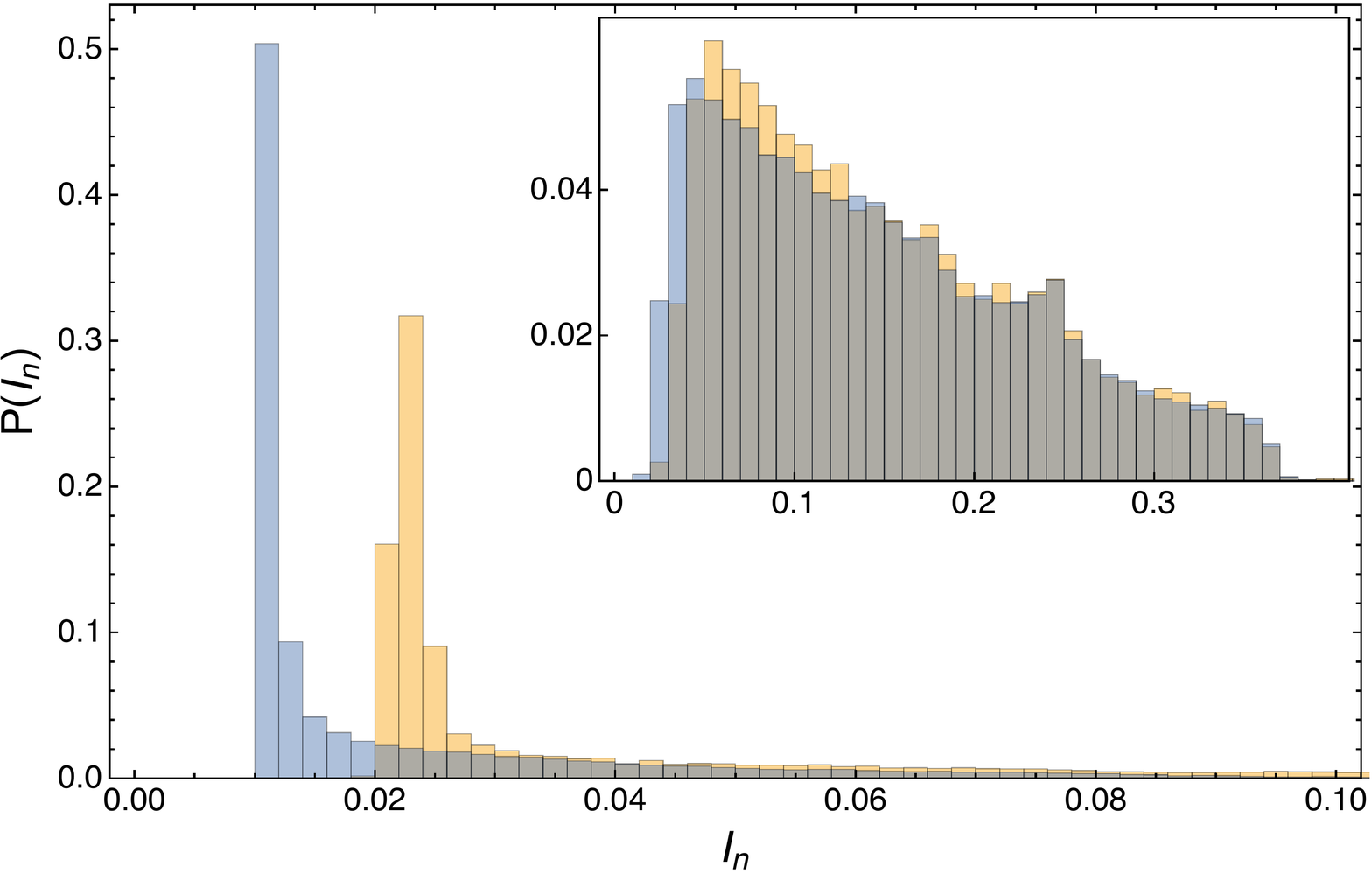}
\caption{Histograms of the inverse participation ratios for
single-particle energy eigenstates for system sizes $L=64$ (yellow)
and $L=128$ (blue) averaged over 1000 disorder realizations with
probability distribution for inhomogeneities given by the box distribution
with $W=3$. Inset: same for eigenstates of \fr{HFF}, \fr{HFF3} 
with $s=0$, $s'=0.2$.}
\label{fig:ipr}
\end{center}
\end{figure}

The particular case $\eta=\pi/2$ maps to non-interacting spinless
fermions by means of a Jordan-Wigner transformation \cite{LSM}. The
resulting Hamiltonian (\ref{Hamil}) is block-diagonal
$H=P_+H_++P_-H_-$, where 
$P_\pm=\frac{1}{2}(1\pm(-1)^{F})$ are projection operators onto the subspaces
with even and odd numbers of fermions respectively ($F$ is the
fermion number operator). We find
\be
H_+=\sum_{j<k=1}^Lc^\dagger_jA_{jk}c_k+{\rm h.c.}\ ,
\label{HFF}
\ee
where  $A_{1,L-1}=2i\tanh(\xi_{L})$,
$A_{1,L}=\frac{2}{\cosh(\xi_{L})}$ and
\be
A_{2j\pm1,2j}=-\frac{2}{\cosh(\xi_{2j})},\
A_{2j-1,2j+1}=2i\tanh(\xi_{2j})\ ,
\label{HFF2}
\ee
Fermions tunnel between neighbouring sites with amplitudes that are
random apart from the constraints $A_{2j-1,2j}=A_{2j,2j+1}$. 
In addition there is a next-nearest neighbour hopping on the
sublattice of all odd sites. Importantly the 
corresponding tunneling amplitudes $A_{2j-1,2j+1}$ are not independent
random variables, but are related to the amplitudes $A_{2j-1,2j}$. The
fermion hopping \fr{HFF2} can therefore be thought of as realizing
a particular kind of \emph{correlated disorder}. As we will see, this
has important consequences for the physical properties of energy
eigenstates. Single-particle energy eigenstates are constructed as
$|\Psi_n\rangle=\sum_{j=1}^L\phi_{n,j}c^\dagger_j|0\rangle$, where  
$\phi_n$ are the (orthonormal) eigenvectors of the matrix $A$ and
$|0\rangle$ is the state without fermions. In order to investigate
whether the model \fr{HFF} is localized we have determined the inverse
participation ratio of single-particle energy eigenstates
$I_n=\sum_{j=1}^L|\phi_{n,j}|^4$. We have considered several probability
distributions of the random parameters $\xi_{2j}$, all of which lead
to the same conclusion. We therefore focus on \fr{PD1}. In
Fig.~\ref{fig:ipr} we show normalized histograms of $I_n$ averaged
over 1000 disorder realizations for $W=3$ and two different system sizes.
We see that the inverse participation ratios are strongly peaked at a
value that we find to scale inversely with system size as $1/L$. This
indicates that the eigenstates are not localized. 
At this point the question arises whether the model \fr{HFF},
\fr{HFF2} is delocalized as a result of fine tuning, or whether it is
representative of a broader class of theories. To investigate this 
issue we have considered free fermion models of the type \fr{HFF} 
with tunneling amplitudes
\bea
A_{2j\pm1,2j}&=&-2|x_{2j}|,\nn
A_{2j-1,2j+1}&=&2is\ \sgn(x_{2j})\sqrt{1-x_{2j}^2}
+s'y_{2j},
\label{HFF3}
\eea
where we take $x_{2j}$ and $y_{2j}$ to be independent random
variables with probability distribution $P_1(x)$ \fr{PD1}. The tuning
parameters $0\leq s,s'\leq 1$ allow us to interpolate between the
``Yang-Baxter'' case in which the next-nearest neighbor tunneling
amplitudes $A_{2j-1,2j+1}$ are fixed in terms of the $A_{2j,2j+1}$ and
the limit in which they become independent random variables. 
We have analyzed IPRs for a range of values $s$ and
$s'$. 
The results
suggest that for $s\approx 1$ and small values of $s'$, i.e. Hamiltonians
close to the Yang-Baxter point, eigenstates are
delocalized. On the other hand for small values of $s$ and $s'$, i.e. weak
uncorrelated next-nearest neighbour tunneling, the data
is more consistent with localization as shown in the inset of
Fig.~\ref{fig:ipr}. This suggests that the ``Yang-Baxter'' model
\fr{HFF}, \fr{HFF2} does not correspond to an isolated point in
parameter space but is representative of a delocalized region that arises
as a result of the correlation between the nearest neighbor and
next-nearest neighbor tunneling.  
\subsection{Local Quantum Quench}
\begin{figure}[ht]
\begin{center}
\includegraphics[width=0.23\textwidth]{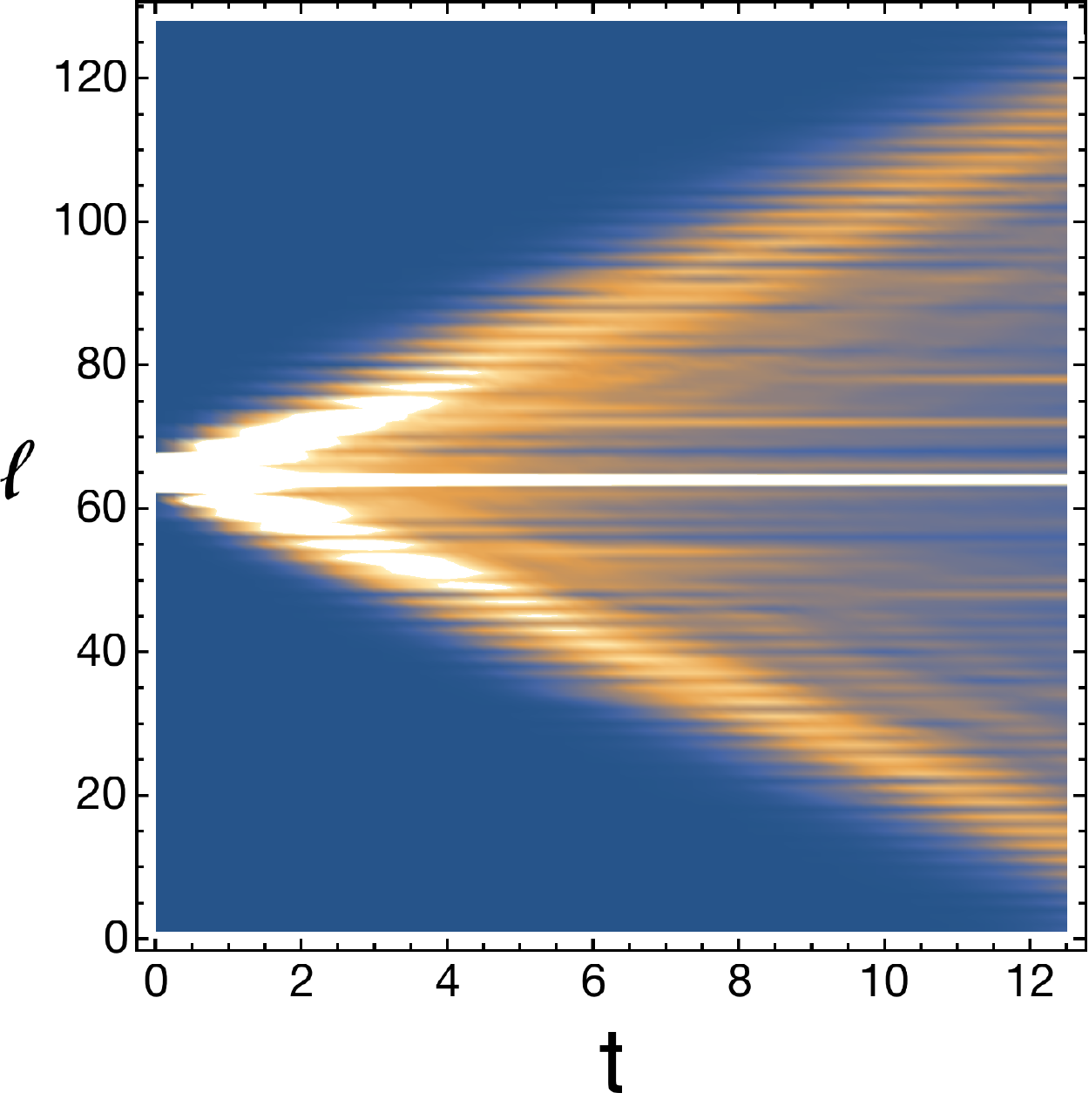}\quad
\includegraphics[width=0.222\textwidth]{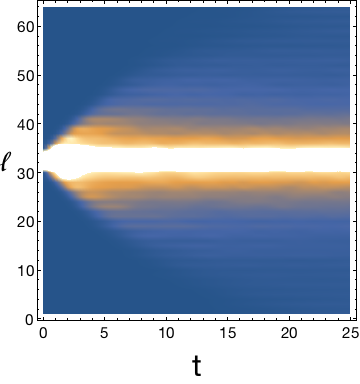}
\caption{
Left plot: $\langle S^z_\ell(t)\rangle$ averaged over $30$ disorder realizations from the box probability distribution for a  system of size $L=128$
and initial thermal state with $\beta=1$. There is a clear
light cone effect. Right plot: the same for the modified free fermion model \fr{HFF3} with $s=s'=0$, $L=64$. Picture is consistent with localization. }  
\label{fig:beta1}
\end{center}
\end{figure}

A second way of investigating localization properties in energy
eigenstates is by considering the spreading of correlations after a
local quantum quench. We prepare the system in the initial finite
energy density state and then overturn two neighboring spins. This
choice of initial state allows us to work in the even fermion
parity sector of the Hilbert space, $(-1)^{\hat{F}}=1$. In order to
investigate the spreading of correlations we determine the expectation
value of the $z$-component of spin at site $\ell$. Using Wick's
theorem we obtain compact expressions for $S^{z}_{\ell}(t)$ that can be
evaluated numerically for systems of hundreds of spins. In
Fig.~(\ref{fig:beta1}) we show results for a representative example,
where a system of size $L=128$ is initially prepared in an energy eigenstate
corresponding to inverse temperature $\beta=1$.
We see that the perturbation, which is initially localized at sites $L/2$ and
$L/2+1$, propagates ballistically through the system, as can be seen
from the presence of a ``light-cone'' outside of which our observable
remains negligibly small. The velocity characterizing this ballistic
propagation depends on the disorder distribution and can be determined
exactly in the thermodynamic limit. The spreading of a local
perturbation in energy eigenstates of the modified free fermion model
\fr{HFF3} can be analyzed in an analogous way. As shown in Fig.~\ref{fig:beta1},
for small values of $s$ and $s'$ the perturbation remains localized at sites
$L/2$ and $L/2+1$ in an extended time window even though a weak
light-cone effect occurs at early times. This again indicates that the
modified free fermion model is localized at small values of $s$, $s'$.

\section{Strongly interacting regime}
\begin{figure}[ht]
\includegraphics[width=0.48\textwidth]{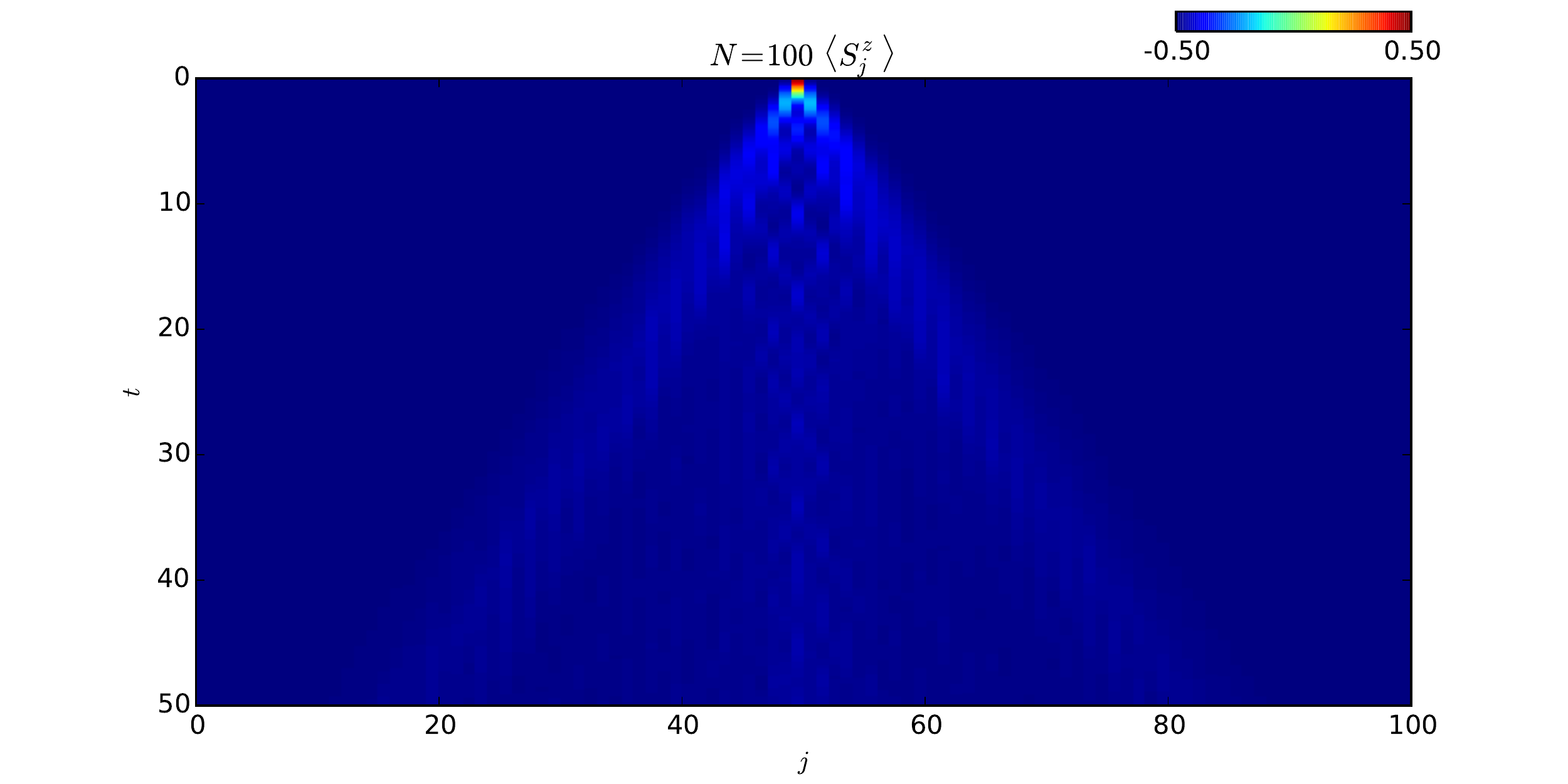}
\caption{Spreading of a single spin flip on top of
the saturated ferromagnetic state for $L=100$ and $\eta=0$ in
(\ref{Hamil}), averaged over 50 disorder realizations with
distribution $P_1(\xi)$. 
}    
\label{fig:quench}
\end{figure}

\begin{figure}[ht]
\includegraphics[width=0.48\textwidth]{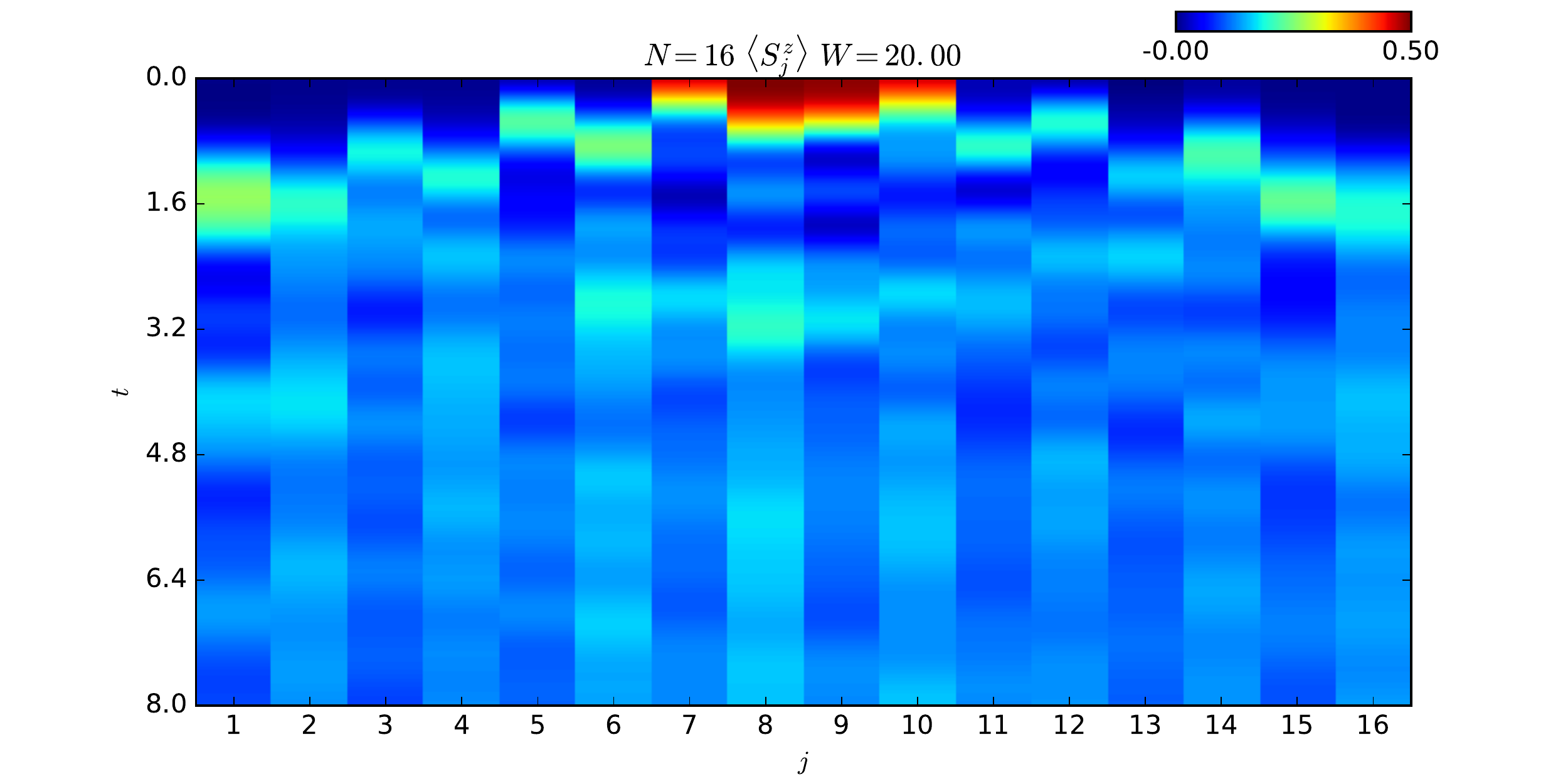}
\caption{(b) Spreading of a spin flips at two
neighbouring sites on top of the ground state for $N=16$ and
$\cos(\eta)=2$ in (\ref{Hamil}), averaged over 20 disorder realizations with
distribution $P_{20}(\xi)$.
}    
\label{fig:quench2}
\end{figure}

Examination of the IPR of \fr{Hamil} away
from the free fermion point for small system sizes $L=10, 12$ is
compatible with delocalized behaviour of energy eigenstates. We also have 
studied the spreading of local perturbations in energy eigenstates.
(i) We have considered a single spin flip at an odd site on top of the
saturated ferromagnetic state. Representative results for the
subsequent dynamics on an $L=100$ site system are shown in
Fig.~(\ref{fig:quench}). There is a clear light-cone effect that
signals ballistic spreading of the perturbation. (ii) We have flipped two
neighbouring spins in the ground state, \emph{cf.}
Ref.~\onlinecite{ganahl} for a discussion of the analogous protocol in
the clean system. In this case however our numerics is limited however
to small systems of up to $L=16$. We find that there
again is a clear light cone effect, see Fig.~\ref{fig:quench2}. 

\section{Bounds on spin and energy transport}
We will now demonstrate that the eigenstates of \fr{Hamil} exhibit
ballistic energy and spin transport for any anisotropy $\eta$ and
disorder strength $W$. We employ a combination of two methods:
the first is based on Mazur's inequality \cite{Mazur} and was
previously employed to establish the existence of a finite temperature
Drude weight in the clean case \cite{Zotos}, while the second is based
on the recently developed hydrodynamic approach to transport in
integrable models \cite{IN,IN2,hydro1,hydro2}. The starting point of
the first approach is the existence of a set of conserved quantities
$[H,{\cal Q}_n]=0$ that are orthogonal in the sense that $\langle
{\cal Q}_n\ {\cal Q}_m\rangle_\beta=\delta_{n,m} \langle {\cal
  Q}_n^2\rangle_\beta.$ Here $\langle . \rangle_\beta$ denotes a
thermal expectation value. As the z-component of total spin
$\sigma^z=\sum_{j=1}^L\sigma_j^z$ is a conserved quantity in our model
we employ a magnetic field term to fix the magnetization in our
thermal ensemble. Given an operator $A=A^\dagger$ with $\langle
A\rangle_\beta=0$ the following inequality due to Mazur \cite{Mazur}
then holds 
\be
\lim_{T_0\to\infty}\frac{1}{T_0}\int_0^{T_0} dt\ \langle
A(t)A\rangle_\beta
\geq \sum_n\frac{\langle A{\cal Q}_{n}\rangle_\beta^2}{\langle
  \big({\cal Q}_{n}\big)^2\rangle_\beta}\ .
\label{Mazur}
\ee
A positive bound for the right-hand side of \fr{Mazur} implies that the
autocorrelation function of the operator $A$ does not decay to zero at
late times. This implies that the Fourier transform has a
non-vanishing (generalized) Drude weight 
\be
\frac{1}{L}\int_0^\infty dt\ \cos(\omega t) \langle
A(t)A\rangle_\beta= 2\pi D_A\delta(\omega)+\dots
\label{Drude}
\ee
When $A$ is the spin current or the energy current
operator the non-decay of the autocorrelation functions shows that
the system is an ideal conductor of spin/energy.
The Hamiltonian \fr{Hamil} has an extensive number of integrals of
motion $Q^{(n)}$ \fr{HCL}. The conservation laws relevant to us here have local
densities and we focus on the most local of these, $Q^{(3)}$, which
involves interactions between spins on at most five neighbouring
sites, \emph{cf.} eqn \fr{Q3}. We furthermore constrain our discussion 
to infinite temperatures $\beta=0$. For local operators the
corresponding thermal average equals the expectation value in
typical energy eigenstates at the associated energy density, which
allows us to draw conclusions about the local properties of the
eigenstates of \fr{Hamil}. In order to use the Mazur inequality \fr{Mazur}
we carry out a subtraction ${\cal Q}_3=Q^{(3)}-\langle Q^{(3)}\rangle_{\beta=0}$,
which ensures that the expectation value of ${\cal Q}_3^2$ is
extensive, i.e. $\lim_{L\to\infty}L^{-1}\langle {\cal
  Q}_3^2\rangle_{\beta=0}=a_1>0.$ The expression for $a_1$ is very
cumbersome so that we do not report it here.

The spin and energy current operators $J^{S,E}$ associated with the
Hamiltonian \fr{Hamil} $H=\sum_jH_{2j-1,2j,2j+1}$ are obtained from
the continuity equations 
\bea
i\sum_{j=-\infty}^\ell[\sigma^z_j,H]&=&J^{\rm S}_\ell,\nn
i\sum_{j=-\infty}^\ell[H_{2j-1,2j,2j+1},H]&=&J^{\rm E}_{2\ell}\ .
\eea
Evaluating the commutators and then summing over all sites gives
\bea
J^{\rm E}&=&4i\sin^2(\eta)\sum_jQ^{(3,3)}_{2j-1,2j,2j+1,2j+2}\nn
&&\qquad\qquad+Q^{(3,4)}_{2j-1,2j,2j+1,2j+2,2j+3} ,
\label{en-curr}
\eea
where $Q^{(3,3)}$ and $Q^{(3,4)}$ are given in \fr{Qnm}.
The spin current operator can be written in the form
\begin{widetext}
\bea
J^{\rm S}&=&2\sum_{j}
J^{(1)}_{2j} \big( T^{xy}_{2j-1,2j}-T^{yx}_{2j-1,2j}
+T^{xy}_{2j,2j+1}-T^{yx}_{2j,2j+1}\big)
+2 J^{(2)}_{2j}\big(T^{xy}_{2j-1,2j+1}-T^{yx}_{2j-1,2j+1}\big)\nn
&-&\frac{2K_{2j}}{\Delta_{2j}}
\big(T^{xzx}_{2j-1,2j,2j+1}+T^{yzy}_{2j-1,2j,2j+1}\big)
+K_{2j}
\big(T^{zxx}_{2j-1,2j,2j+1}+T^{zyy}_{2j-1,2j,2j+1}
+T^{yyz}_{2j-1,2j,2j+1}+T^{xxz}_{2j-1,2j,2j+1}\big),
\label{spin-curr}
\eea
\end{widetext}
where we have defined
\be
T^{\alpha_1\dots\alpha_n}_{j_1,\dots,j_n}=\prod_{k=1}^n\sigma^{\alpha_k}_{j_k}\ .
\ee
We find that in contrast to the homogeneous case, the energy current is not
conserved, i.e. $[H,J^E]\neq 0$.

At infinite
temperature and finite magnetization ${\rm m}$ a tedious but
straightforward calculation gives the following result for the overlap
of the spin current with the third conserved charge
\be
\langle J^{\rm S} {\cal Q}_3\rangle_{\beta=0}=
\frac{  \Delta}{1-\Delta^2}
\sum_{n} \frac{4 \text{m} \left(1-4
  \text{m}^2\right)f(\xi_{2n})}{[\cosh (2 \xi_{2n})-(\cos (2
  \eta))]^3}\ ,
\ee
where
\bea
f(z)&=&\cos(2\eta)\cosh(6z)\nn
&-&2\big(\cos(4\eta)-\cos (2 \eta )+3\big) \cosh (4z)
\nn
&+&\big(6\cos(4 \eta )-\cos (6 \eta )+10\big) \cosh (2 z)\nn
&-&18 \cos (2 \eta )+2 \cos(4\eta )+6\ .
\eea
For a very large system we may replace the sum by an integral so that
\bea
&&\langle J^{\rm S} {\cal Q}_3\rangle_{\beta=0}= a_{\rm S}L+o(L)\ ,\nn
&&a_{\rm S}=\frac{\Delta}{1-\Delta^2}
\int d\xi \frac{4 \text{m} \left(1-4 \text{m}^2\right)f(\xi)P(\xi)}{\big(\cosh (2 \xi)-\cos (2
  \eta)\big)^3}.
\label{JQ}
\eea
Here $P(\xi)$ is the probability distribution on the random variables
$\xi_{2n}$. Importantly we have $a_{\rm S}\neq 0$ unless we fine-tune
the probability distribution. This in turn provides a positive bound
for the Mazur inequality
\be
\lim_{L \to\infty}
\lim_{T_0\to\infty}\frac{1}{T_0L}\int_0^{T_0} dt\ \langle
J^{\rm S}(t)J^{\rm S}\rangle_{\beta=0}
\geq \frac{a_{\rm{S}}^{2}}{a_1}\ .
\ee
In the case of the energy current for simplicity we consider the zero
magnetization sector ${\rm m}=0$. Applying Mazur's inequality we find
\bea
&&\lim_{L \to\infty}
\lim_{T_0\to\infty}\frac{1}{T_0L}\int_0^{T_0} dt\ \langle
J^{\rm E}(t)J^{\rm E}\rangle_{\beta=0}\nn
&&\geq \lim_{L\to\infty}\frac{1}{L}\frac{\langle J^{\rm E}{\cal
    Q}_3\rangle_{\beta=0}^2}{\langle {\cal Q}_3^2\rangle_{\beta=0}}
=\frac{64\big(2+2\cos(2\eta)\big)^2}{16\sin^4(\eta)a_1}\ .
\label{Ebound}
\eea
Interestingly the bound \fr{Ebound} is independent of the
inhomogeneities. The generalization to ${\rm m}\neq 0$ is
very tedious but straightforward and provides a non-zero bound as
well. 

The above calculation proves that at energy densities corresponding to
infinite temperature the model \fr{Hamil} exhibits (i) a non-zero 
Drude weight at any finite magnetization; (ii) ballistic energy transport.

\section{Spin and energy transport from generalized hydrodynamics}

Generalized Drude weights \fr{Drude} can be analyzed in full by means of
the approach introduced in Ref.~\onlinecite{IN}. The starting point is the
existence of a basis of local charges $\hat{Q}_i$ and associated
currents $J_i$. Using these charges a {\it generalized Gibbs ensemble}
is defined by the density matrix $\rho_{\rm GGE}\sim \exp(-\sum_{n}\mu_{n}\hat{Q}_{n})$,
where $\mu_{i}$ are ``chemical potentials''. The generalized Drude
weights $D_A$ are then obtained from appropriate expectation values in
this ensemble and are determined by using the thermodynamic Bethe
ansatz (TBA) method \cite{takahashi}. According to Ref.~\onlinecite{IN},
in integrable models $D_A$ can be expressed as
\be
D_A=\sum_{n}\int d\lambda
\frac{\eta_n(\lambda)}{\rho_n^{\rm tot}(\lambda)}
\left(\frac{\epsilon'_{n}(\lambda)q^{\rm
    eff}_{A}(\lambda)}{2\pi(1+\eta_n(\lambda))}
\right)^2 ,
\label{D}
\ee
where $\eta_n(\lambda)=\bar{\rho}_{n}(\lambda)/\rho_{n}(\lambda)$ is
the ratio of hole and particle densities, 
$\rho^{\rm  tot}_{n}(\lambda,\{\xi_{2j}\})=\rho_{n}+\bar{\rho}_{n}$,
$\epsilon_{n}(\lambda)$ are the energies of n-string excitations over
the state of thermal equilibrium \cite{BEL} and 
$q^{\rm eff}_{A}=\partial_{\mu_{A}}\log\eta_{n}$ are
effective transport charges. The implementation of this approach in
our ``inhomogeneous'' case reveals (see Appendix \ref{app:hydro} for
more details) that the disorder merely renormalizes the Drude weight
through the disorder-dependence of the velocity of the elementary
excitations over the equilibrium state under consideration, which
enters \fr{Drude} via the factor $1/\rho^{\rm tot}_n(\lambda)$. It
follows then that the disorder average can be exchanged with the
integration and summation in (\ref{D}). The disorder averaged Drude
weight is then given by 
\be
\overline{D}_A\!=\!\sum_{n}\int d\lambda
\overline{[\rho_n^{\rm tot}(\lambda)]^{-1}}
\eta_n(\lambda)
\left[\frac{\epsilon'_{n}(\lambda)q^{\rm
    eff}_{A}(\lambda)}{2\pi(1+\eta_n(\lambda))}
\right]^2,
\ee
where $\overline{[\rho_n^{\rm tot}(\lambda)]^{-1}}=
\int
P(\{\xi\})\frac{1}{\rho_{n}^{\rm tot}(\lambda,\{\xi\})}$
denotes the disorder average with probability distribution function $P(\xi)$.
As the total density $\rho^{tot}_{n}(\lambda)$ is a positive
quantity this average is non-zero for generic $P(\xi)$. Therefore,
the Drude weight is only renormalized due to the disorder dependence
of string particle and hole densities. We note that in contrast to the
Mazur bound calculation the TBA approach takes into account the full
set of conserved quantities. These observations can be universally
extended to any integrable model with disorder of the type described here.

\section{Conclusions}
In this paper we studied a Yang-Baxter integrable interacting spin
system with controllable short-range correlated disorder. Using
a combination of diagnostics we have demonstrated the absence of
many-body localization. 
We find that the model is in fact an ideal
conductor for both energy and magnetization. For particular parameter
values the model can be mapped to non-interacting fermions and we have
established the absence of Anderson localization in this case. In
contrast, a sufficiently strong deformation of the free-fermion Hamiltonian away
from the Yang-Baxter point shows signatures of localization. We expect
that in the interacting case small perturbations away from the
Yang-Baxter point will lead to diffusive behaviour , while
sufficiently strong deformations will be required to induce an MBL
transition.
\acknowledgments
We are grateful to M. Brockmann, J.-S. Caux and E. Ilievski for
collaboration in the early stages of this project. We thank
W. Buijsman, A. de Luca, A. Pal, S. Parameswaran and V. Yudson for
very helpful discussions. This work was supported by the EPSRC under
grant EP/N01930X (FHLE) and the Delta-ITP consortium (VG), a program
of the Netherlands Organization for Scientific Research funded by the
Dutch Ministry of Education, Culture and Science.

\appendix
\section{Inhomogeneous XXZ chain}
\label{app:inhom}
The Quantum Inverse Scattering Method  (QISM)\cite{Korepinbook} provides a
simple way of introducing ``impurities'' into Yang-Baxter integrable
models. This has been used in the literature to construct a variety of
models with impurities embedded in both non-interacting and correlated
hosts \cite{AJ,imp1,imp2,imp3,imp4,imp5,imp6,imp7,imp8,imp9}, as well
as models with ``disorder'' \cite{KZ1,KZ2,Z1,Z11,Z2}. 
Here we focus on the simplest case, which is related to the spin-1/2 Heisenberg XXZ
chain. The basic ingredients in the QISM are the R-matrix 
$R(\mu)\in {\text End}(V_A\otimes V_A)$ and the L-operator
$L(\mu)\in {\text End}(V_A\otimes V_Q)$, where $V_A$ and $V_Q$ are
finite-dimensional ``auxiliary'' and ``quantum'' vector spaces.
In the cases we are interested in the Yang-Baxter relations read
\be
R(\lambda-\mu)\left[L(\lambda)\otimes L(\mu)\right]
=\left[L(\mu)\otimes L(\lambda)\right]R(\lambda-\mu).
\label{YBE}
\ee
In the case of the spin-1/2 XXZ chain we have \cite{Korepinbook}
\bea
\big(L(\lambda)\big)^{ab}_{\alpha\beta}&=&\frac{1+\tau^z_{ab}\sigma^z_{\alpha\beta}}{2}
+b(\lambda)\frac{1-\tau^z_{ab}\sigma^z_{\alpha\beta}}{2}\nn
&&
+c(\lambda)\big(\tau^-_{ab}\sigma^+_{\alpha\beta}+\tau^+_{ab}\sigma^-_{\alpha\beta}\big)\ ,\nn
b(\lambda)&=&\frac{\sinh(\lambda)}{\sinh(\lambda+i\eta)}\ ,\
c(\lambda)=\frac{i\sin(\eta)}{\sinh(\lambda+i\eta)}\ ,
\label{Lop}
\eea
where $\eta$ is a free parameter and $\tau^\alpha$, $\sigma^\alpha$
are Pauli matrices acting on the auxiliary and quantum spaces
respectively. The QISM provides a commuting family of transfer
matrices $[\tau(\mu),\tau(\lambda)]=0$ of the form
\bea
\tau(\mu)_{\alpha_1,\dots,\alpha_L}^{\beta_1,\dots,\beta_L}
&=&\prod_{j=1}^L\Big[L(\mu-\xi_j)\Big]_{\alpha_j\beta_j}^{c_jc_{j+1}}\ ,
\eea
where the free parameters $\xi_j$ are known as ``inhomogeneities'' and
where we have defined $c_{L+1}=c_1$. In
order to obtain a local Hamiltonian we now set
\be
\xi_{2j+1}=0\ ,
\ee
and then take the logarithmic derivative of the transfer matrix at $\mu=0$
\be
H=2i\sin\eta\frac{d}{d\mu}\Big|_{\mu=0}\ln\big(\tau(\mu)\big).
\ee
The explicit expression for the resulting Hamiltonian is given by \fr{Hamil}.

\subsection{Spectral properties}
The Hamiltonian \fr{Hamil} is readily diagonalized by Algebraic Bethe
Ansatz \cite{Korepinbook}. The energy eigenvalues are given by
\be
E=-\sum_{j=1}^N\frac{4\sin^2(\eta)}{\cosh(2\lambda_j)-\cos(\eta)}\ ,
\label{EXXZ}
\ee
where the rapidities $\lambda_1,\dots,\lambda_N$ are solutions of
the Bethe Ansatz equations 
\bea
&&\left(\frac{\sinh(\lambda_j+i\eta/2)}
{\sinh(\lambda_j-i\eta/2)}\right)^{\frac{L}{2}}\ 
\prod_{k=1}^{L/2}\frac{\sinh(\lambda_j-\xi_{2k}+i\eta/2)}
     {\sinh(\lambda_j-\xi_{2k}-i\eta/2)}\nn
&&     =\prod_{k\neq j}
\frac{\sinh(\lambda_j-\lambda_k+i\eta)}
{\sinh(\lambda_j-\lambda_k-i\eta)}\ ,\quad
j=1,\dots, N.
\label{BAE}
\eea
Equations \fr{EXXZ} and \fr{BAE} establish a peculiar property of the
model \fr{Hamil}: the spectrum is invariant under arbitrary
permutations of the inhomogeneities
$\{\xi_2,\xi_4,\dots,\xi_{L}\}$, i.e.
\be
{\rm spec}\ H\big[\{\xi_{2},\dots,\xi_L\}\big]=
{\rm spec}\ H\big[\{\xi_{P(2)},\dots,\xi_{P(L)}\}\big]
\label{specsymm}
\ee
for any permutation $P$ of the integers $2,4,\dots, L$. This property
is not apparent from the explicit expression \fr{Hamil} and Hamiltonians
corresponding to different permutations of the inhomogeneities
generally do not commute.

\subsubsection{Free Fermion Point}
The Hamiltonian \fr{Hamil} has a free fermion point at 
$\eta=\frac{\pi}{2}$. The corresponding Hamiltonian is
\bea
H&=&\sum_{j=1}^{L/2}\frac{1}{\cosh(\xi_{2j})}\sum_{\alpha=x,y}
\left[\sigma^\alpha_{2j-1}\sigma^\alpha_{2j}+\sigma^\alpha_{2j}\sigma^\alpha_{2j+1}
\right]\nn
  &-&\sum_{j=1}^{L/2}\tanh(\xi_{2j})
\left[\sigma^y_{2j-1}\sigma^z_{2j}\sigma^x_{2j+1}-
\sigma^x_{2j-1}\sigma^z_{2j}\sigma^y_{2j+1}\right].\nn
\label{HXX}
\eea
By applying the Jordan-Wigner transformation on can bring the
Eq. (\ref{HXX}) into the form of the Eq. 

\subsubsection{Isotropic (XXX) Limit}
The SU(2) invariant versions of the Hamiltonian and the Bethe Ansatz
equations are obtained by redefining
\be
\lambda_j=\frac{\eta}{2}\Lambda_j\ ,\quad
\xi_{2j}=\frac{\eta}{2}\gamma_j\ ,
\ee
and then taking the limit $\eta\to 0$. This gives a Hamiltonian of the
form 
\bea
H&=&\sum_j\frac{4}{\gamma_j^2+4}\left[\vec\sigma_{2j-1}\cdot\vec\sigma_{2j}
  +\vec\sigma_{2j}\cdot\vec\sigma_{2j+1}-2\right]\nn
&-&\sum_j\frac{2\gamma_j}{\gamma_j^2+4}\vec{\sigma}_{2j}\cdot
\big(\vec{\sigma}_{2j-1}\times\vec{\sigma}_{2j+1}\big) \nn
&+&\sum_j\frac{\gamma_j^2}{\gamma_j^2+4}
\big(\vec\sigma_{2j-1}\cdot\vec\sigma_{2j+1}-1\big).
\eea
The Bethe Ansatz equations become
\be
\left(\frac{\Lambda_j-i}{\Lambda_j+i}\right)^{\frac{M}{2}}
\prod_{k=1}^{M/2}\frac{\Lambda_j-\gamma_k-i}{\Lambda_j-\gamma_k+i}
=\prod_{k\neq j}
\frac{\Lambda_j-\Lambda_k-2i}
{\Lambda_j-\Lambda_k+2i}\ .
\label{BAEXXX}
\ee
The energy corresponding to a solution of \fr{BAEXXX} is
\be
E=-\sum_j\frac{8}{\Lambda_j^2+1}\ .
\ee

\section{Drude weights from the TBA calculations}
\label{app:hydro}

Let us consider first the case of $|\Delta|>1$. While any eigenstate in a
finite system of size $L$ is assigned a unique set of rapidities $\{\lambda_{j}\}_{k=1}^{N}$ taken from solutions of Bethe equations (\ref{BAE}), in the thermodynamic limit (defined as $L\rightarrow\infty$, $N\rightarrow\infty$ with $N/L$ finite), the solutions to Bethe equations organize into regular patters which indicate the
presence of well-defined particle excitations. These correspond to
magnons and their bound states,so-called Bethe strings
\cite{takahashi}. A general string solution reads $\{\lambda_{\alpha}^{k,m}\}=\{\lambda_{\alpha}^{k}+(k+1-2m)\frac{i\eta}{2}\}$, where $m=1,2,\ldots, k$ and $\alpha$ numerates different $k$-strings and $m$ runs over internal rapidities. Scattering of different magnonic particles are characterized by the amplitudes 
\bea
S_{j}&=&\frac{\sin(\lambda-j\frac{i\eta}{2})}{\sin(\lambda+j\frac{i\eta}{2}},\nonumber\\
S_{jk}&=&\prod_{m=-\frac{k-1}{2}}^{\frac{k-1}{2}}\prod_{n=-\frac{j-1}{2}}^{\frac{j-1}{2}}S_{2m+2n+2}\nonumber\\
&=&S_{|j-k|}S_{j+k}\prod_{m=1}^{\mbox{min}(j,k)-1}S^{2}_{|j-k|+2m}
\label{s-mat}
\eea
with convention that $S_{0}\equiv 1$. In the thermodynamic limit  particle rapidities become densely distributed along the real axis in the rapidity plane. This permits to introduce distributions $\rho_{k}(\lambda)$ of $k$-string particles, along with the dual hole distributions $\bar{\rho}_{k}(\lambda)$ (holes are solutions to Bethe ansatz equations  which differ from Bethe roots $\lambda_{k}$). The discrete Bethe equations (\ref{BAE}) get replaced by the integral Bethe-Yang equations. Assuming validity of the string solution in the presence of $M$ inhomogeneities ($M/N\leq 1/2$), we can write these integral equations for the densities of string particles and holes in the thermodynamic limit of the inhomogeneous case.
The Bethe-Yang equations for particles $\rho_{n}(\lambda)$ and holes $\bar{\rho}_{n}(\lambda)$ are given by
\bea
& &\frac{1}{N}\left(\sum_{j=1}^{M}a_{n}(\lambda+\xi_{j})+(N-M)a_{n}(\lambda)\right)\nonumber\\
&=&\bar{\rho}_{n}(\lambda)+A_{nm}\star \rho_{m}(\lambda).
\label{TBA1}
\eea   
Here, the explicit form of the functions $a_{n}(\lambda)$ and
$A_{nm}(\lambda) = \delta_{nm}\delta(\lambda)+a_{nm}$, which depend on
the anisotropy parameter $\Delta$, can be obtained from the following
relations  
\bea
a_{n}(\lambda)&=&\frac{1}{2\pi i}\partial_{\lambda}\log S_{n}(\lambda), \nonumber\\ a_{nm}(\lambda)&=&\frac{1}{2\pi i} \partial_{\lambda}\log S_{nm}(\lambda)
\label{kern}
\eea
where indexes $n,m$ label corresponding stringy content. The $\star$
operation refers to the convolution with the kernel $A_{nm}$  
\bea
A_{mn}\star\rho_{m}(x)\equiv \sum_{m}\int_{-Q}^{Q}dy A_{mn}(x-y)\rho_{m}(y)
\eea
where the integration and summation limits depend on the value of anisotropy parameter. Explicitly, for $\Delta>1$ we have 
\bea
a_{n}(\lambda)=\frac{1}{2\pi}\frac{\eta\sinh(n\eta)}{\cosh(n\eta)-\cos(\eta\lambda)}.
\eea
For the isotropic (XXX) situation, when $\eta\rightarrow 0$, the driving function and the kernel are given by
\bea
a_{n}(\lambda)&=&\frac{1}{\pi}\frac{n}{(n^{2})+\lambda^{2}},\\
A_{nm}(\lambda)&=&\delta(\lambda)\delta_{nm}+(1-\delta_{nm})a_{|n-m|}(\lambda)\\
&+&2 a_{|n-m|+2}(\lambda)+\ldots +2a_{n+m-2}(\lambda)+a_{n+m}(\lambda)\nonumber
\eea
while in this case $Q=\infty$ and sum runs to infinity as well.

Classification of the particle content in the gapless regime $|\Delta|<1$ is more involved, details can be found in \cite{TS}, \cite{takahashi}. Here, in addition to the magnon
type label $k$, an extra parity label $v\in\pm$ is required. Importantly, integers $k$ now no longer coincide with the length of a string, i.e. a number of magnons forming a bound state. Instead, the $k$-th particle consists of $n_{k}$ Bethe roots and carries parity $v_{k}$ (see \cite{TS} for further details). Setting $\Delta=\cos(\gamma)$, where $\gamma/\pi =m/l$ (with $m$,$l$ co-prime integers)
is a root of unity, the number of distinct particles in the spectrum is finite. Changing the parametrization $\lambda\rightarrow i\lambda$, $\eta\rightarrow i\gamma$  and incorporating the additional parity label, the elementary scattering amplitudes and kernels read
\bea
S_{k}(\lambda)\rightarrow S_{(n_{j},v_{j})}=\frac{\sinh[\lambda-n_{j}\frac{i\gamma}{2}+(1-v_{j})\frac{i\pi}{4}]}{\sinh[u+n_{j}\frac{i\gamma}{2}+(1-v_{j})\frac{i\pi}{4}]}
\eea
and the whole set of scattering kernels is obtained, as in the case of $\Delta >1$ , from Eqs.~(\ref{s-mat}, \ref{kern}). The Bethe-Yang equations gets modified,
\bea
a_{j}(x)=\mbox{sign}(q_{j})(\rho_{j}+ \bar{\rho}_{j})+a_{jk}\star\rho_{k}
\eea
where the summation in the convolution expression runs from $1$ to $m_{l}$ defined as $m_{0}=0$, $m_{i}=\sum_{k=1}^{i}\nu_{k}$ and numbers $\nu_{1},\ldots \nu_{l-1}\geq 1$, $\nu_{l}\geq 2$ participate in the continuum fraction expression for $\gamma/\pi$, e.g. $\gamma/\pi =1/(\nu_{1}+1/(\nu_{2}+1/(\nu_{3}+\ldots)))$. Numbers $q_{j}$ are defined recursively as \cite{takahashi}, $q_{0}=\pi/\gamma$ and 
\bea
q_{j}&=&\frac{1}{2}[(1-\delta_{m_{i},j})q_{j-1}+q_{j+1}],\quad m_{i}\leq j\leq m_{i+1}-2\nonumber\\
q_{j}&=&(1-2\delta_{m_{i-1},j})q_{j-1}+q_{j+1}, \quad j=m_{i}-1, i<l\nonumber
\eea
Explicitly, the kernels $a_{j}(\lambda)$ are given by 
\bea
a_{j}(\lambda)=\frac{1}{2\pi}\frac{\gamma \sin(\gamma q_{j})}{\cosh(\gamma\lambda)+\cos(\gamma q_{j})}
\eea

The most important thing to notice here is that the left hand side (driving terms) of the Bethe-Yang equations depends on the inhomogeneities while the right hand side (convolution kernel) {\it does not} depend on inhomogeneities. This can also be checked by explicit re-derivation of steps leading to these equations (\ref{TBA1}). 

The second set of equations is derived using the variation of the free
energy (per particle) $f=e-Ts$ with respect to $\rho_{n}$ and
$\bar{\rho}_{n}$. Here 
\be
e=2\pi\sum_{n}a_{n}(\lambda)\rho_{n}(\lambda)
\ee
is the energy density and the entropy density is
\bea
s&=&\sum_{n=1}^{\infty}\int_{-\infty}^{\infty} d\lambda
\Big[(\rho_{n}+\bar{\rho}_{n})\ln(\rho_{n}+\bar{\rho}_{n})\nn 
&&-\rho_{n}\ln\rho_{n}-\bar{\rho}_{n}\ln\bar{\rho}_{n}\Big].
\eea
Variation of (\ref{TBA1}) leads to the relationship between $\delta\rho_{n}$ and $\delta\bar{\rho}_{n}$,
\bea
\delta\bar{\rho}_{n}=-A_{nm}\star \delta\rho_{n}.
\eea
which finally leads to the second TBA equation
\bea
\ln(1+\eta_{n})=\frac{2\pi J }{T}a_{n}
+A_{nm}\star \ln(1+\eta_m^{-1})\ ,
\label{TBA2}
\eea
where $\eta_{n}=\bar{\rho}_{n}/\rho_{n}$. Importantly, since the right
hand side of (\ref{TBA1}) does not depend on $\delta\rho_{n}$
or on $\delta\bar{\rho}_{n}$, eqn (\ref{TBA2}) and hence
$\eta_n$ is {\it independent of the inhomogeneities}. It is customary
to re-cast \fr{TBA2} in terms of the dressed energies defined by
$\varepsilon_n=T\ln(\eta_n)$ 
\be
\frac{\varepsilon_j}{T}=\frac{\varepsilon_j^{(0)}}{T}
+a_{nm}\star \ln(1+e^{-\varepsilon_m/T}),
\ee
where the bare energies are $\varepsilon_j^{(0)}=2\pi J a_{n}$.
In \cite{IN} a hydrodynamic approach to the Drude weight(s) has been
formulated based on the TBA approach.  The starting
point is the existence of a basis of local charges $\hat{Q}_i$ and
associated currents $J_i$. Using these charges a {\it generalized Gibbs
ensemble} is defined by the density matrix
\bea
\rho_{\rm GGE}\sim \exp(-\sum_{n}\mu_{n}\hat{Q}_{n}),
\eea
where $\mu_{i}$ are ``chemical potentials''. The generalized Drude
weights $D_A$ are then obtained from appropriate expectation values in
this ensemble and are determined by using the thermodynamic Bethe
ansatz (TBA) method \cite{takahashi}. According to Ref.~\cite{IN}
in integrable models $D_A$ can be expressed as
\bea
D_A=\sum_{n}\int d\lambda
\frac{\eta_n(\lambda)}{\rho_n^{\rm tot}(\lambda)}
\left(\frac{\epsilon'_{n}(\lambda)q^{\rm
    eff}_{A}(\lambda)}{2\pi(1+\eta_n(\lambda))}
\right)^2 ,
\label{Drude_SM}
\eea
where 
\be
q^{\rm eff}_{A}=\partial_{\mu_{A}}\log\eta_{n}
\ee
are effective transport charges. The functions $\epsilon'_{n}$ are
derivatives of the energies of elementary excitations over the state
of thermal equilibrium and were calculated in Ref~\cite{BEL}. They are
obtained from the dressed energies by solving a set of linear
integral equations
\bea
\epsilon'_j*(1-K)_{jk}(\lambda)&=&\frac{d\varepsilon_k^{(0)}(\lambda)}{d\lambda}, \\
K_{jk}(x,y)&=&-\sgn(q_j)a_{jk}(x-y)\big(1+e^{\varepsilon_j/T}\big)^{-1}.\nonumber
\eea
The only quantities in \fr{Drude_SM} that depend
on the inhomogeneities $\xi_{2j}$ are the total densities $\rho_n^{\rm
  tot}(\lambda)$. This can be seen from (\ref{TBA1}) once the
disorder-independent equations (\ref{TBA2}) for $\eta_{n}$ have been
solved. It follows that the disorder averaging of the generalized
Drude weights can be interchanged with the integration and summation
in (\ref{Drude_SM}). Introducing
\bea
\overline{\left(\frac{1}{\rho^{\rm tot}_{n}(\lambda)}\right)}=\int
P(\{\xi\})\frac{1}{\rho_{n}^{\rm tot}(\lambda,\{\xi\})}
\eea
where $P(\xi)$ is a disorder probability distribution we then
can express disorder averaged Drude weights in the form
\bea
\overline{D}_A=\sum_{n}\int d\lambda
\overline{\left(\frac{1}{\rho^{\rm tot}_{n}(\lambda)}\right)}
\eta_n(\lambda)
\left(\frac{\epsilon'_{n}(\lambda)q^{\rm
    eff}_{A}(\lambda)}{2\pi(1+\eta_n(\lambda))}
\right)^2 .\nonumber
\eea
As the total density $\rho^{\rm tot}_{n}(\lambda)$ is a positive
quantity $D_A$ is only renormalized due to the dependence of the
string particle and string hole densities on disorder, and will not vanish
unless the disorder probability distribution is fine-tuned.


\end{document}